# A Preliminary Survey of Semantic Descriptive Model for Images

Chengxi Yan
School of Information Resource Management
Renmin University of China
Beijing China
20218113@ruc.edu.cn

Jie Jian
School of Information Resource Management
Renmin University of China
Beijing China
2023202344@ruc.edu.cn

Yang Li
Chinese Academy of Science and Education Evaluation
Hangzhou Dianzi University
Hangzhou China
liyang0815@hdu.edu.cn

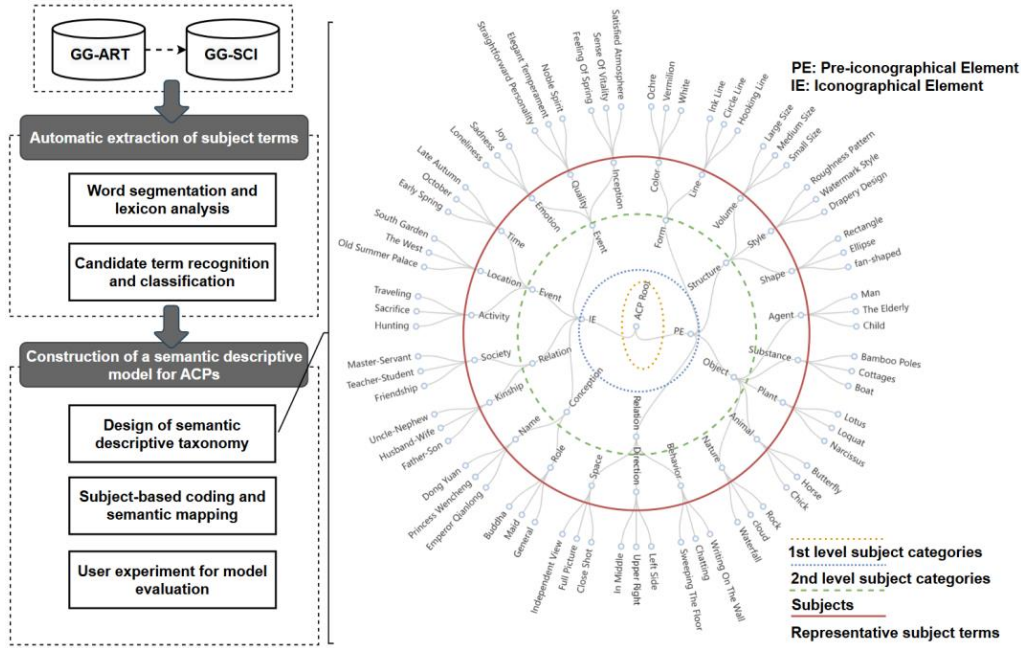

**Figure 1: The research procedure and taxonomy of semantic descriptive model (only shown with three examples of subject terms)**

## ABSTRACT

Considering the lack of a unified framework for image description and deep cultural analysis at the subject level in the field of Ancient Chinese Paintings (ACP), this study utilized the Beijing Palace Museum's ACP collections to develop a semantic model integrating the iconological theory with a new workflow for term extraction and mapping. Our findings underscore the model's effectiveness. SDM can be used to support further art-related knowledge organization and cultural exploration of ACPs.

• Computing methodologies → Modeling and simulation.

## 1 Introduction

Painting is not only a manifestation of human creativity and intelligence, but also can reflect the philosophical and cultural contexts of specific historical periods. At present, many museums have transformed ancient paintings into digital images, making it easier for everyone to appreciate them. However, the current descriptions of these paintings tend to focus mainly on basic information, such as the name of a painting or an artist, without delving into the subjects and stories expressed within the artworks. This may prevent viewers from fully understanding the significance of the paintings while appreciating them. Worse still, the field of ACP has long lacked a universal semantic descriptive specification which led to the inability to fully reveal the artistic connotations and cultural values, thereby having a negative impact on the development of art image analysis. This raised an challengeable research question: How do we use a quantitative approach to automatically extract subject terms and construct a user-friendly semantic descriptive model for ACPs that can comprehensively reflect the structure of subject knowledge, making them more useful for domain research?

## 2 Research Design

We proposed a two-step analytical solution (Figure 1). We collected ACP-related resources from the Beijing Palace Museum (BPM, one of the most famous national museums), including titles, images, descriptions, and keywords, forming the "GG-ART" dataset including 1600 paintings spanning 1600 years. To expand subject terms, we retrieved ACP-related documents from the CNKI database (the most common-used Chinese database for academic journals) based on GG-ART titles. Experts verified document types and filtering conditions (20 influential journals, each with ≥5 relevant documents). We obtained 3041 scientific documents, forming "GG-SCI," with details like title, author, abstract, and keywords.

*Automatic extraction of subject terms*. A two-stage automatic approach was designed for term extraction. The first stage was coarse-grained term recognition, which required the word segmentation of sentences in the abstracts of both datasets and a series of lexicon analysis-based processing involving sentence segmentation, part-of-speech (POS) tagging, stopwords-based removal and POS filtering for nouns and adjective. The chunking of noun phrases was then performed based on relevant arrangement pattern rules (e.g. "nouns", "adjectives+nouns" or "nouns+nouns") [1]. In the second stage, term ranking was conducted to distinguish those terms with great importance. The score function can be designed by combining the contextual embedding of sentences and the similarity between terms. Here we used an advanced deep learning-based model "EmbedRank" to calculate the embedding [2]. In addition, we merge extracted candidate terms with keywords and clustered them based on the representation similarity using the K-means algorithm, which resulted in multiple subject clusters with similar semantic terms.

*Construction of a semantic descriptive model for ACPs*. Based on Panofsky's iconological theory [3], we designed a Semantic Descriptive Model (SDM) with a three-layer structure (shown in Figure 1). The first layer includes "pre-iconographical elements" (i.e. PE, namely the basic information describing the natural subject matters of images) and "iconographical element" (i.e. IE, namely the conventional interpretations of the objects in images). In the second and third layer, SDM can provide more specific subject categories/subjects of PE and IE, such as "color" and "line" under the category of "form". Moreover, the obtained subject clusters were mapped to these SDM subjects through semantic matching, which relies on those experts' manual alignment and adjustment (e.g. to manually move a subject term in specific cluster to a target subject). During the construction of the model, three domain experts were engaged in the repeated discussion, encoding and modification of SDM, to ensure the accuracy of SDM. A consistency test based on the kappa statistics [4] for the expert coding results for SDM were carried out to validate the reliability of the model. Also, to verify the SDM's effectiveness, a user evaluation test based on our designed online interface (Figure 2) was created, in which 8 professional scholars were required to assess the usefulness when operating with/without SDM (i.e. the baseline system only utilized information provided by BPM) for randomly-selected 100 ACPs. Inspired by the USE questionnaire [5], We designed a similar questionnaire with a 5-point Likert scale to measure diversity, comprehension, effectiveness, and satisfaction. The score ranged from 1 (strongly disagree) to 5 (strongly agree), and 3 indicated that viewers think the usefulness of the two systems is the same.

## 3 Result Analysis

We reported the results of the evaluation test (shown in Figure 2). Clearly for each question, the mean value of users' evaluative scores is statistically significantly higher than 3 points, showing SDM outperforms the baseline in various aspects of usefulness. We can fully understand the advantages of SDM, and foresee its potential value in artistic analysis and relevant research (see Q3).

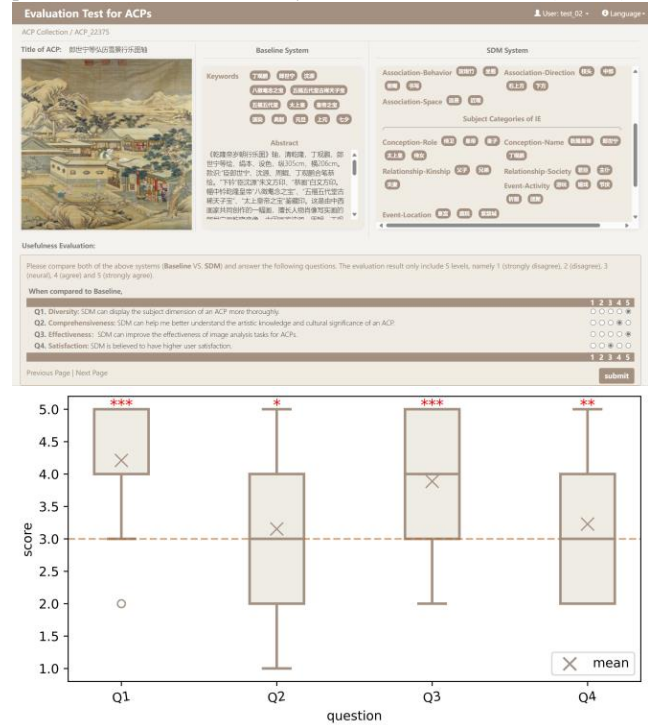

**Figure 2: user interface and evaluative result with the one-sample t-test (*, **, *** respectively denote p<0.05, 0.01, 0.001)**

## 4 Conclusion

In this paper, a hierarchical semantic description model "SDM" was developed for ACPs in the semi-automatic way. The empirical results testified the advantage of SDM. This preliminary research is believed to provide new organization approach and analytical perspectives to the field of Chinese painting arts.

## ACKNOWLEDGMENTS
This project is supported by the grant from National Natural Science Foundation of China (NO. 72204258).